\documentclass{article}
\usepackage{ijcai19}
\usepackage{times}
\usepackage{soul}
\usepackage{url}
\usepackage[hidelinks]{hyperref}
\usepackage[utf8]{inputenc}
\usepackage{graphicx}
\usepackage{amsmath}
\usepackage{booktabs}
\usepackage{algorithm}
\usepackage{algorithmic}
\usepackage{amssymb}
\usepackage{booktabs} 
\usepackage{tabu}
\usepackage{adjustbox}
\usepackage{multirow}
\usepackage{comment}

\usepackage{comment}

\usepackage{booktabs}

\usepackage[shortlabels]{enumitem}

\usepackage{caption}
\DeclareCaptionLabelFormat{andtable}{#1~#2  \&  \tablename~\thetable}

\newcommand{\bE}{\mathbb{E}}
\newcommand{\bR}{\mathbb{R}}
\newcommand{\cA}{\mathcal{A}}
\newcommand{\cM}{\mathcal{M}}
\newcommand{\cS}{\mathcal{S}}
\newcommand{\cT}{\mathcal{T}}
\newcommand{\cO}{\mathcal{O}}
\newcommand{\cU}{\mathcal{U}}

\usepackage{bbm}
\usepackage{color}
\title{A Regulation Enforcement Solution for Multi-agent Reinforcement Learning}

\author{
Fan-Yun Sun$^1$
\and
Yen-Yu Chang$^1$\and
Yueh-Hua Wu$^{1,2}$\And
Shou-De Lin$^1$
\affiliations
$^1$National Taiwan University\\
$^2$Riken-AIP\\
\emails
\{b04902045,b03901138\}@ntu.edu.tw,
\{d06922005, sdlin\}@csie.ntu.edu.tw
}

\begin{document}

\maketitle

\begin{abstract}


Human behaviors are regularized by a variety of norms or regulations, either to maintain orders or to enhance social welfare. If artificially intelligent (AI) agents make decisions on behalf of human beings, we would hope they can also follow established regulations while interacting with humans or other AI agents. However, it is possible that an AI agent can opt to disobey the regulations (being defective) for self-interests. In this paper, we aim to answer the following question: \textbf{Consider a multi-agent decentralized environment. Agents make decisions in complete isolation of other agents. Each agent knows the state of its own MDP and its own actions but it does not know the states and the actions taken by other players. There are a set of regulations for all agents to follow. Although most agents are benign and will comply to regulations but not all agents are compliant at first, can we develop a framework such that it is in the self-interest of non-compliant agents to comply after all?} We formulate the problem using reinforcement learning and game theory and then propose a solution based on the key idea that although we could not alter how defective agents choose to behave, we can, however, leverage the aggregated power of compliant agents to boycott the defective ones. We conducted simulated experiments on two scenarios: \emph{Replenishing Resource Management Dilemma} and \emph{Diminishing Reward Shaping Enforcement}, using deep multi-agent reinforcement learning algorithms. We further use empirical game-theoretic analysis to show that the method alters the resulting empirical payoff matrices in a way that promotes compliance (making mutual compliant a Nash Equilibrium).

\end{abstract}

\section{Introduction}
Consider the driving matrix game shown below.
\begin{table}[h]
\centering
\begin{tabular}{c|c|c|c|}
{\small Driving Matrix Game}  & $L$ & $R$ \\
\hline
$L$ & $1, 1$ & $0, 0$ \\
\hline
$R$ & $0, 0$ & $1, 1$ \\
\hline
\end{tabular}
\end{table}

Two autonomous cars driving on a road against each other, they have to choose either to swerve on the left ($L$) or to swerve on the right ($R$) of the road. 
Introducing regulations (e.g. the right car should yield) here can mitigate ambiguities and avoid lose-lose situations. Furthermore, some regulations are not there to prevent agents from making malevolent decisions or to enhance agents' self-interest, they are there for ethical reasons or to enhance welfare of the society as a whole. Human beings opt to follow such regulations even if it undermines their self-interests because either they are afraid of being punished by authorities (i.e. government), or they are well-educated with civic consciousness. Unfortunately, such awareness may not exist for some AI agents that are trained to maximize individual rewards. In other words, without certain special design (e.g hard-coding ethical rules for agents to follow or specifically trained toward altruism), we shall not expect a normal AI agent to obey regulations that lead to sacrifice of its rewards. Similar to human society, even a small amount of AI agents not compliant to existing regulations can lead to catastrophe. 

Consider a real-world dilemma - \emph{Replenishing Resource Management Dilemma}. It describes a situation in which group members share a replenishing resource (e.g. lumbering or fishing) that will continue to produce benefits unless being over-harvested~\cite{wiki:Social_dilemma}. Regulations such as \emph{International Convention for the Regulation of Whaling} are signed by many countries to constrain the harvesting behavior. In the future, it is likely that robots become the main force to harvest such resources, and thus it is crucial to design a framework to prevent agents from violating the regulation to maximize self-interests.

There have been some works aiming at designing \emph{ethical} AI agent instead of one that only optimizes its own rewards. For example, assuming in a multi-agent environment, \cite{sun2018designing} proposes a design for benevolent (non-greedy) agents through shaping the reward function. They propose the idea of \emph{diminished rewards} that leads to less satisfaction for consecutive rewards, and consequently achieves non-greediness of agents as they are not motivated to obtain resources rapidly and repeatedly. In the experiment consisting of both stronger and weaker agents, it is shown that implementing such reward function can lead to more balanced distribution of resources. 
Although the diminishing reward function seems to be a favorable solution from the social-welfare point of view, there is no incentive for the stronger agents to implement such feature since it hurts their overall rewards. Enforcing \emph{every single agent} to comply can be difficult.

 
In this paper, we aim to address the following problem, named \textbf{Regulation Enforcement}: There are regulations that the society expect all agents to comply, but certain individuals can gain advantage by not complying. 
Our goal is to design a solution such that it is in the self-interest of non-compliant agents to comply. There are many literature that discuss how to regulate agents behaviour with many mental constructs such as BDI agent research \cite{meneguzzi2009norm,alechina2012programming,lee2014n}.
However, it is not applicable in our scenario as we consider the scenario where no centralized authority can control agents' internal design. Penalizing \textit{Defective}~\footnote{We name agents that comply to all regulations as \textit{Compliant} and agents that disobey one or more regulations as \textit{Defective}.} agents is only possible when there are special purpose agents deployed (i.e. ``police'' as in real world) but this is not our case.

 
 

 We make the assumption that at least a certain percentage of agents are \textit{Compliant}, which is reasonable in most real-world scenarios. Our solution leverages the power of the crowd, eliminating the need of deploying special purpose agents. There are two major components: a detector and a boycotting strategy. The boycotting strategy states that agents shall shape their policy in a way that boycotts the non-complying agents, which is identified by the detector. By implementing this framework, we expect \textit{Defective} agents to lose incentives to disobey regulations since being \textit{Defective} can increase the chance of being detected and then boycotted, which results in lower return. 



We summarize our contributions as below:
\begin{itemize}
    \item To our knowledge, this is the first work to introduce the task of \textbf{Regulation Enforcement}. We believe it could become a crucial problem with the pervasiveness of AI agents. We further provide a formal definition from aspects of reinforcement learning and game theory. Note that we consider different settings when compared with not literature of normative agent-based systems.
    
    \item We propose a simple yet effective solution to solve this problem in a decentralized environment. Our solution contains a detector and a general boycotting policy. Although we could not directly alter the policy of \textit{Defective} agents in a decentralized environment, the \textit{Compliant} agents are highly motivated to comply since then they can contribute to the prevention of \textit{Defective} agents by lessening their rewards.
    
    \item We evaluate the effectiveness of our model on simulated scenarios of \emph{Replenishing Resource Management Dilemma} and \emph{Diminishing Reward Shaping Enforcement}. We also use \emph{empirical game-theoretic analysis} to further show how empirical payoff matrices evolve after applying our method. Results shown are promising. 
\end{itemize}
\section{Preliminaries}

\subsection{Normative Agent-Based Systems} In normative systems, regulations are enforced in regimentations or sanctions. Regimentation explicitly prevents agents from reaching certain states of affairs (i.e. close a road to prevent cars from entering the road) while sanctions impose punishments on agents that violate the norms (i.e.fines for traffic violation). However, neither of this will work under our scenario because there are no authorities that have the power to change the environment, constrain agents’ behavior, or impose punishments on individuals. In our paper, we are leveraging the collective power of compliant agents to sanction those defective ones. This is a major difference when comparing our work with related works on normative multi-agent systems.

\subsection{Reinforcement Learning}
RL defines a class of algorithms solving problems modeled as a Markov Decision Process (MDP). A $n$-player partially observable Markov game $\cM$ is defined by a set of states $\cS$ and an observation function $O: \cS \times \{1,2, ...,n\} \rightarrow \bR^d$ specifying each player's $d$-dimensional view, along with $n$ sets of actions allowable from any state $\{\cA_1, \cA_2, ..., \cA_n\}$, one for each player, a transition function $\cT: \cS \times \cA_1 \times \cA_2 \times ... \times \cA_n \rightarrow \Delta(\cS)$, where $\Delta(\cS)$ denotes the set of discrete probability distributions over $\cS$, and a reward function for each player $i$: $r_i: \cS \times \cA_1 \times \cA_2 \times ... \times \cA_n \rightarrow \bR$.
Let $\cO_i = \{ o_i~|~s \in \cS, o_i = O(s,i)\}$ be the observation space of player $i$, to choose actions, each player uses policy $\pi_i : \cO_i \rightarrow \Delta(\cA_i)$.

For temporal discount factor $\gamma \in [0, 1]$ we can define the long-term payoff as $V^{\vec{\pi}}_i(s_0)$ for player $i$ when the joint policy $\vec{\pi} = (\pi_1, \pi_2, ..., \pi_n)$ is followed starting from state $s_0 \in \cS$. 

\begin{equation} \label{eq:expected_payoff}
V^{\vec{\pi}}_i(s_0) = \bE_{\vec{a}_t \sim \vec{\pi}(O(s_t)), s_{t+1} \sim \cT(s_t, \vec{a}_t)}\left[ \sum_{t = 0}^\infty \gamma^t r_i(s_t, \vec{a}_t) \right].
\end{equation}

\subsection{Game Theory}

A \textbf{normal-form game} is a tuple $(S, f, n)$ where $n$ is the number of players, $S_i$ is the strategy set for player $i$, $S=S_1 \times S_2 \times \dotsb \times S_n$ is the set of strategy profiles and $f(x)=(f_1(x), \dotsc, f_n(x))$ is its payoff function evaluated at $x \in S$. Let $x_i$ be a strategy of player $i$ and $x_{-i}$ be a strategy profile of all players excluding player $i$. When each player $i \in \{1, \dotsc, n\}$ chooses strategy $x_i$ resulting in strategy profile $x = (x_1, \dotsc, x_n)$ then player $i$ obtains payoff $f_i(x)$. Note that the payoff depends on the strategy profile chosen, i.e., on the strategy chosen by player $i$ as well as the strategies chosen by all the other players. \textbf{Extensive-form
games} extend these formalisms to the multistep sequential case (e.g. poker).

\textbf{Matrix Games} are two-player games where each player has two strategies to choose from. It is the special case of two-player perfectly observable ($O_i(s) = s$) Markov games obtained when $|\cS| = 1$ and $\cA_1 = \cA_2 = \{C, D\}$, where $C$ and $D$ are called (atomic) cooperate and defect respectively. Matrix Games serve as mathematical model of many of the simplest conflict situation in the areas of economics, mathematical statistics, war science, and biology.

A \textbf{Nash Equilibrium} is a strategy profile $x^* \in S$ such that no unilateral deviation in strategy by any single player is profitable for that player, that is,
\begin{equation} \label{eq:nash}
    \forall i,x_i\in S_i :  f_i(x^*_{i}, x^*_{-i}) \geq f_i(x_{i},x^*_{-i}).
\end{equation}
When the inequality above holds strictly (with $>$ instead of $\ge$) for all players and all feasible alternative strategies, the equilibrium is classified as a \textit{Strict Nash Equilibrium}. If instead, for some player, there is exact equality between $x^*_i$ and some other strategy in the set $S$, then the equilibrium is classified as a \textit{Weak Nash Equilibrium}. 

\textbf{Empirical game-theoretic analysis} (EGTA) is the study of meta-strategies (or styles of play) obtained through simulation in complex games \cite{walsh2002analyzing}.

\begin{figure*}[h]
\centering
\includegraphics[width=.8\linewidth]{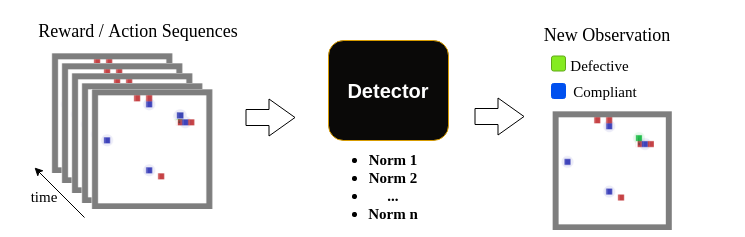}
\caption{A simple illustration of the proposed framework. Red blocks denote the shared resource while blue blocks denote agents. The detector takes a sequence of past actions or rewards as input and determines which agents are \textit{Defective}. Agents then incorporate that information into their observation and take action according to their policies trained with Boycotting Reward Shaping (Equation \ref{eq:boycott}).}
\label{fig:framework}

\end{figure*}

\section{Problem Formulation}
Our goal is to design a strategy to discourage the violation of regulations to gain more rewards in multi-agent scenarios where agents make their own decisions. Note that we use the term \textit{regulation} instead of \textit{constraint} because agents can possibly violate regulation but not constraint. Our targeted problem is different from constrained MDPs \cite{altman1999constrained}. The proposed problem, namely the \textbf{Regulation Enforcement}, is formulated as below:

Let $\cM$ be a $n$-player Markov game and there are $N$ regulations that regularize the agents' behavior. Regulations can be defined in two ways:
\begin{itemize}
    \item Defined in the \textbf{reward function space} such as requiring agents to shape their reward function in a certain way, i.e. implement diminishing reward shaping \cite{sun2018designing} so that resource is distributed more equally among agents.
    \item Defined in the \textbf{policy space} like requiring agents to behave in a certain way in specific situations, i.e. stopping the car at a red light can be formulated as $\pi(\text{state = observe red light}) = \text{stop}$.
\end{itemize}
$\Pi^C$ denotes the set of policies that follow all regulations, and $\Pi^D$ denotes the set of policies that violate one or more regulations. In this paper, agent $i$ with the policy $\pi_i$ is labelled as \textit{Compliant(C)} if $\pi_i \in \Pi^C$ and \textit{Defective(D)} if $\pi_i \in \Pi^D$. 
Under this setting, $\Pi^C \bigcap \Pi^D = \emptyset$ and $\Pi^C \bigcup \Pi^D = \Pi$ (the set of all legal policies). 

We denote the set $(\pi_1, \pi_2, ..., \pi_n)$ as the resulting joint policy under the assumption that at least $M\%$ (M=80 in our experiments) of agents are \textit{Compliant} ($\pi_j \in \Pi^C$). Let $\pi^{C}_j$, $\pi^{D}_j$ denote the resulting policy of agent $j$ being \textit{Compliant} and \textit{Defective} respectively. The demand of \emph{Regulation Enforcement} comes from the following assumption: 
\begin{equation} \label{eq:eq3}
\exists j \text{ s.t. }
V_j^{(\pi_1, \pi_2, ..., \pi^{C}_j, ... \pi_n)}(s_0) < V_j^{(\pi_1, \pi_2, ..., \pi^{D}_j, ... \pi_n)}(s_0)
\end{equation}
where $s_0$ is the starting state. 
That means there exists some agents who can gain more rewards by being \textit{Defective}.
Note that we could not alter the behaviors of \textit{Defective} agents since agents make decisions in a decentralized manner, but we can affect \textit{Compliant} agents' policies in a way that lessens the return of  \textit{Defective} agent $j$, $V_j^{(\pi_1, \pi_2, ..., \pi^{D}_j, ... \pi_n)}(s_0)$. That is, our goal is to design a strategy for agents (in particular \textit{Compliant} agents) so that being \textit{Defective} can damage the overall reward:
\begin{equation} \label{eq:eq4}
\forall j, 
V_j^{(\pi_1, \pi_2, ..., \pi^{C}_j, ... \pi_n)}(s_0) \geq V_j^{(\pi_1, \pi_2, ..., \pi^{D}_j, ... \pi_n)}(s_0)
\end{equation}
As mentioned in the preliminaries, general-sum matrix games is the special case of two-player perfectly observable Markov game when there is only one state and both players have only two strategies to choose from. Similarly, if we take the case where player $i$ has only two strategies $\{C_i, D_i\}$ to choose from $\forall i$, then the problem can be rephrased from the point-of-view of game theory, as described below.

Let $(S, f, n)$ be a normal-form game with $n$ players, where $S_i$ is the set of strategy for player $i$, $S=S_1 \times S_2 \times \dotsb \times S_n$ is the set of strategy profiles and $f(x)=(f_1(x), \dotsc, f_n(x))$ is its payoff function evaluated at $x \in S$. Given that the strategy set for player $i$ can be denoted as $\{C_i, D_i\}$, the set of strategy profiles can be denoted as $\{C_1, D_1\} \times \{C_2, D_2\} \times ... \times \{C_n, D_n\}$. 
Let the strategy that player $i$ takes as $s_i$,
and $x$ be any strategy profile that consists of at least $M \%$ ($M =80$ in our experiments) of \textit{Compliant} strategies, then Equation \eqref{eq:eq3} becomes: 
\begin{equation}
    \exists i \text{ s.t. }  f_i(C_i, x^*_{-i}) < f_i(D_i,x^*_{-i}).
\end{equation}
The goal of \emph{Regulation Enforcement} then becomes:  
\begin{equation} \label{eq:goal}
    \forall i:  f_i(C_i, x^*_{-i}) \geq f_i(D_i,x^*_{-i}).
\end{equation}
where $x_i$ is a strategy profile of player $i$ and $x_{-i}$ is a strategy profile of all players excluding player $i$. Note that we adopt similar notation as in Equation \eqref{eq:nash}.

This is a realistic assumption. In real world, regulators (i.e. governments) tend to lay down legislation but not all individuals will be compliant. Furthermore, regulators often don’t have any idea who is going to be compliant at first, they often use detectors (i.e. speed cameras) to detect defective individuals.


\section{Proposed Framework}
Intuitively, we aim to mitigate agents' incentive to disobey regulations. The goal is to lessen the rewards gained being \textit{Defective} comparing to those gained being \textit{Compliant}. If this can be achieved, then any rational agents will choose to be \textit{Compliant}. 
 Note that in a decentralized environment, we cannot force any agent to implement or execute any strategy because each agent makes its own decision.
Thus, our plan is to offer a framework that can benefit an agent in the long run, so it has high motivation to implement and execute the compliant strategy. 
The framework states that if defecting agents are detected, an agent should shape its reward function towards boycotting them, as illustrated in Figure \ref{fig:framework}. Note that an assumption is made: at least $M$\% of players are \textit{Compliant} ($M$ has to represent the majority). Furthermore, since all agents interact with one another in an environment with shared resources, it is assumed that they can observe how many rewards (resources) other agents have collected. Intuitively, the proposed method is trying to boycott \textit{Defective} agents by leveraging the aggregated power of \textit{Compliant} agents. 

There are two major components in our method: training a detector and laying down a boycott strategy.
\subsection{Detector}
This detector makes prediction of \textit{Defective} agents by observing agents' behavior. More specifically, it takes reward sequences and/or action sequences (if needed) of an agent as input and learns to classify whether the agent is \textit{Compliant} or \textit{Defective}. The underlining hypothesis is that since the goal of a \textit{Defective} agent is to obtain more rewards through not obeying regulations, \textit{Defective} agents shall be detectable based on the their actions performed and sequence of rewards obtained. More formally, let $\vec{A}_{i,t}$ denote the sequence of actions and/or rewards of agent $i$ up till time $t$, we aim to learn a detector $\mathcal{D}(\vec{A}_{i,t}, \theta)$ parameterized by $\theta$ that outputs 1 (True) if agent $i$ is classified as \textit{Defective} or 0 (False) if agent $i$ is classified as \textit{Compliant}. Multiple inferences can be made at a time. 
In many scenarios, a rule-based detector is sufficient. Take the \emph{Replenishing Resource Management Dilemma} for instance, one simple rule is sufficient to determine whether a resource-gathering agent exceeds the maximal quota allowed. However, some scenarios can be less trivial and a more sophisticated classifier is required for detection. For example, to detect whether a comity function is implemented in an auto-driving agent.

In reality, the detector can be trained either centrally or in a distributed manner. For example in human society, there are detectors trained and deployed globally (i.e. surveillance cameras) yet there are also detectors that varies among individuals (i.e. moral conscience). In our experiments, the detector is trained as a global behavior detector for simplicity.

\subsection{Boycotting Reward Shaping}


We exploit the idea of Reward shaping \cite{ng1999policy} to design the boycott strategy. Reward shaping is initially proposed as an efficient way of including prior knowledge in the learning problems so as to enhance the convergence rate. Additional intermediate rewards are provided to enrich a sparse base reward signal, giving the agent with useful gradient information. 

In \cite{sun2018designing}, instead of using reward shaping as a way of enhancing convergence rate, they use reward shaping to shape agents' policies in an intended way. They suggest designing a benevolent agent based on a reward shaping method which diminishes rewards to make the agent feel less satisfied for consecutive rewards. 
\begin{align} \label{eq:dim1}
\mathcal{R}^{'}(s_t, a_t, \mathcal{I}_t)= \mathcal{R}(s_t, a_t) \times\mathcal{F}(\mathcal{I}_t)
\end{align}
\begin{align} \label{eq:dim2}
\mathcal{I}_t = \sum_{i=1}^{\mathcal{W}}\mathcal{R}(s_{t-i}, a_{t-i})
\end{align}
$\mathcal{F}$ is a predetermined non-strictly decreasing function and $\mathcal{W}$ is a chosen window size.

Similar to \cite{sun2018designing}, we use reward shaping as a method of shaping agents' resulting policies. 
The idea states that agents should optimize a \textit{mental-reward} that is usually different from the actual rewards obtained. We plan to design a reward shaping scheme that encourages agents to boycott \textit{Defective} agents while maximizing their own reward. More formally, \textbf{Boycotting Reward Shaping} is defined below:

Denote the trained detector as $\mathcal{D}$ where $\mathcal{D}_{t}(i)$ outputs 1 if it classifies agent $i$ as \textit{Defective} or 0 if it classifies agent $i$ as \textit{Compliant}. Let the reward function of agent $i$ be $\mathcal{R}_i^{'}(s_t,a_t)$, and the number of agents be $N$, agents have to optimize a reward function $\mathcal{R}_i^{'}(s_t,a_t)$ which is defined as
\begin{align} \label{eq:boycott}
    \mathcal{R}_i^{'}(s_t,a_t) = \mathcal{R}_i(s_t,a_t) - B \times \frac{[\sum_{j = 1}^{N}\mathcal{D}_{t}(j) \times \mathcal{R}^{\text{obs}}_j(s_t,a_t)]}{\sum_{j=1}^{N}\mathcal{D}_{t}(j)}
\end{align}
where $B$ is a predetermined ratio which we refer to as the \textit{Boycotting Ratio}. Note that agents are not aware of the value functions of other agents but they can observe the amount of resources consumed by other agents in an environment of shared utility. We use  $\mathcal{R}^{\text{obs}}_j$ to denote the reward of agent $j$ observed by agent $i$. The rightmost term denotes the average ``observed'' reward of all \textit{Defective} agents. Note that $B=0$ corresponds to the original scenario where no changes are applied. 

\begin{figure*}[t]
\centering
\includegraphics[width=.6\linewidth]{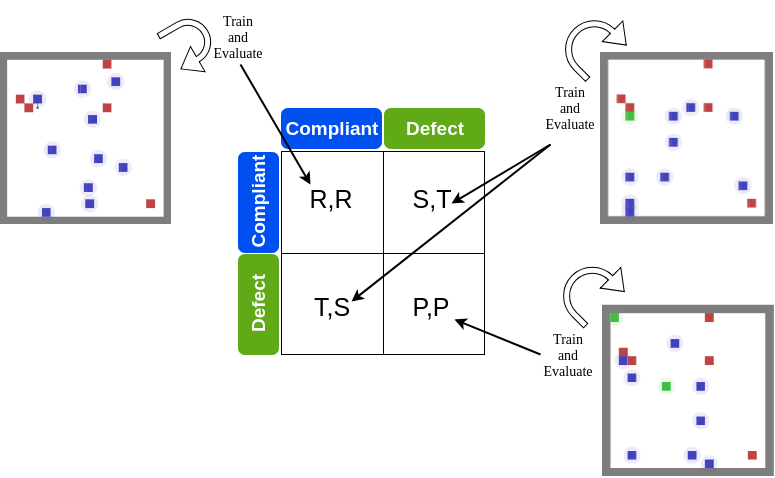}
\caption{Workflow to obtain empirical payoff matrices in Experiment 3. Agents that follow all regulations are \textit{Compliant} (blue) otherwise \textit{Defective} (green). There are 10 agents in total and we fix 8 out of 10 agents as \textit{Compliant}. We aim to observe what is the payoff of the other 2 agents when they choose to behave compliantly or defectively. For each entry in the payoff matrix, we train and evaluate the game correspondingly (notice how the number of green defectors vary between entries).  
By repeatedly playing out the games using resulting joint policies, and averaging the results, we can obtain the payoffs for each cell of the matrices. 
\label{fig:tournament}}
\end{figure*}

\section{Experiments}
We conduct three experiments based on deep multi-agent reinforcement learning. In the first two experiments, we address the scenarios of \emph{Replenishing Resource Management Dilemma} and \emph{Diminishing Reward Shaping Enforcement} as mentioned in the introduction section. In the third experiment, we adopt similar settings as  Experiment 1 and use empirical game-theoretic analysis to observe how the proposed method affects the empirical payoff matrices. We make the assumption that at least $M$ percentage of agents abide by all regulation. Motivated by paper of similar line of research~\cite{turner2006designing}, we set $M=80\%$ in our experiments. Note that if the proposed framework is able to solve the problem at $M=80\%$, it will also solve problems where the initial percentage of compliant agents are $ \geq 80\%$.

If we consider human society, starting with $80\%$ of the agents being compliant is a fairly low bound (in other words, we are considering a society where 1 out of 5 people will disobey the law). In our simulation experiments, in order to ensure compliant agents behave so, we chose a random $M\%$ subset of all agents and implement their value functions according to the regulations.

\subsection{Experiment 1: Replenishing Resource Management Dilemma}
In this experiment, we aim to address the scenario where group members share a renewable resource and sustainable development can only be achieved if no individual over harvest the resource. As a result, a regulation is laid down to prevent gaining self-interest from over-harvesting.

To conduct the experiment, we design the following game. There are 5 agents that interact with each other in a 20 x 20 grid world. Apple trees, which appear as red blocks on the map, represent the replenishing resource. An apple tree will die out (disappear) if more than 5 apples are collected, and a new apple tree will appear at a random location on the map. 
The regulation is that agents shall not collect more than 3 apples at any time. However, agents can obviously benefit from not obeying the rule and collecting more than 3 apples at a time. 

This experiment has the following setting:
\begin{itemize}
    \item \textbf{REGULATION: For the sake of sustainable development, all agents shall not collect more than 3 apples at any time.}
    \item \textit{Compliant Agents}: Agents that are not collecting more than 3 apples at any given time. There are 4 \textit{Compliant} agents.
    \item \textit{Defective Agents}: Agents that collect more than 3 apples at one or more times in the past. There is 1 \textit{Defective} agents that collects up to 5 apples at a time.
\end{itemize}
 The percentage of \textit{Compliant} agents $M$ is set to 80\% (4 out of 5 agents are \textit{Compliant}). Note that a rule-based detector $\mathcal{D}$ that examines the collection rate is sufficient in this case. Thus, in this experiment we will focus on the effectiveness of boycotting. 

\subsection{Experiment 2: Diminishing Reward Shaping Enforcement}
In this experiment, we aim to address the scenario where agents are not equally capable and have to share a kind of resource. Since members have varying capabilities, to prevent stronger agents leaving weaker agents ``starving'', the regulation demands every agent to implement and conduct the \emph{diminishing reward function} to act non-greedily as described previously. 

To conduct the experiment, we design the following game. As in Experiment 1, there are 5 agents in a 20 x 20 grid world, 4 out of 5 agents are \textit{Compliant}, and apple trees represent the shared resource. Different from the previous experiment, now the agents are not equally capable: it takes one time step for a stronger agent to collect 5 apples while it takes two time steps for weaker agents to collect 5 apples. The regulation states that all agents are required to implement diminishing reward shaping so they do not act greedily. Understanding that an agent (in particular the stronger one) can obviously benefit from not obeying the rule to behave greedily, here we assume the one particular strong agent to be \textit{Defective} through not implementing the diminishing reward. The goal of this experiment is to evaluate whether the proposed \emph{Regulation Enforcement} framework can find and penalize the \textit{Defective} agent.


This experiment has the following settings:
\begin{itemize}
    \item \textbf{REGULATIONS: For the sake of avoid acting greedily, all agents shall optimize a new reward function $\mathcal{R}^{'}(s_t, a_t, \mathcal{I}_t)$ which is defined as (according to \cite{sun2018designing})} 

    \begin{align}
    \mathcal{R}^{'}(s_t, a_t, \mathcal{I}_t) = \begin{cases}    
     \mathcal{R}(s_t, a_t) & \mathcal{I}_t \leq \tau \\
    -1 & \mathcal{I}_t > \tau
    \end{cases}
`   \end{align}
   \begin{align}
    \mathcal{I}_t = \sum_{i=1}^{3}\mathcal{R}(s_{t-i}, a_{t-i})
    \end{align}
    \textbf{where $\tau$ is set to 2.}

    \item \textit{Compliant}: Agents that implement the diminishing reward function accordingly to the regulation above. There are 4 \textit{Compliant} agents.
    \item \textit{Defective}: Agents that do not implement the diminishing reward function. There is 1 \textit{Defective} agents here.
\end{itemize}

Note that a rule-based detector $\mathcal{D}$ is not sufficient in this case. A binary classifier needs to be trained to decide whether an agent is \textit{Compliant} or not. 
We will evaluate the detection accuracy as well as the effectiveness of the boycotting framework.

\begin{figure}[h]
   \centering
    \includegraphics[width=.65\linewidth]{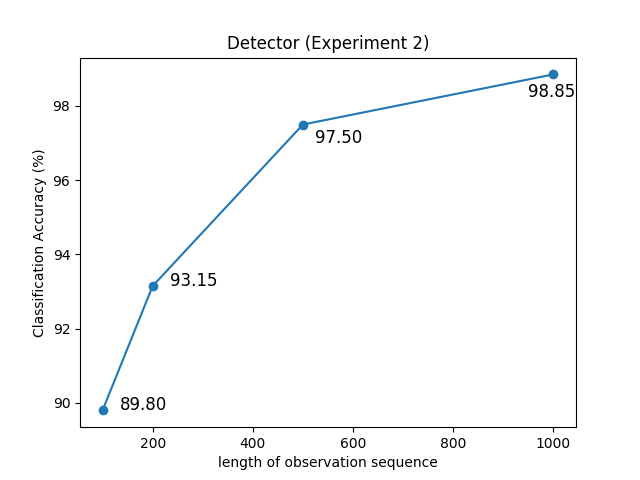}
    \captionof{figure}{Detector accuracy with different length of observation sequence (Experiment 2).}
    \label{fig:exp-detector}
\end{figure}

\subsection{Experiment 3: Payoff Matrices}
In this experiment, we investigate the proposed task and solution using empirical game-theoretic analysis. We aim to observe how empirical payoff matrices evolve before and after applying the regulation enforcement framework. We focus on two players and regard all other 8 agents as part of the dynamic environment. We adopt the scenario of Experiment 1, except that there are now 10 agents instead of 5. We will evaluate on different choice of policies of these two players and make the other 8 agents always \textit{Compliant}. We train and evaluate correspondingly to each situation to obtain the matrices, filling one cell at a time, illustrated in Figure \ref{fig:tournament}.
We set the \textit{Boycotting Ratio} to 2 in this experiment.

\subsection{Simulation Details}
Games studied here are implemented in a large-scale 2D gridworld platform MAgent \cite{zheng2017magent}. The state $s_t$ and the joint action of all players $\vec{a}$ determines the state at the next time-step $s_{t+1}$. Observations of agents depend on the agent's current position and consist of two parts, spatial local view and non-spacial feature. Spatial view consists of several rectangular channels, which includes map of locations of other agents and map of non-penetrable wall. These channels will be masked by a circle and the radius of the circle is defined as $view\_range$. In all our experiments, $view\_range$ is set to 2, which means that the size of one channel is $5 \times 5$, where $2 \times 2+1=5$. Non-spatial feature includes last action, last reward, absolute position of all other agents and apples, normalized position of the agent, and ID embedding. ID embedding is the binary representation of agent's unique ID.
Actions are discrete actions such as move or gather. Similar to the observations, move range and gather range are circular range with their radii denoted as $move\_range$ and $gather\_range$ respectively. In our experiments, we set $move\_range$ to 3 and $gather\_range$ to 1. That makes 33 valid actions in total. Each episode lasts for $1,000$ steps and all results are obtained from an average of 100 episodes after training 30000 episodes. 

We use Double Dueling DQN \cite{van2016deep,wang2015dueling,mnih2015human} to simulate the game since it converges faster than DRQN\cite{hausknecht2015deep} and A2C\cite{mnih2016asynchronous} in our experiment. Default neural networks have two convolution layers both with a $3 \times 3$ kernel and two fully connected dense layer. The spatial view observation is fed into the two convolution layers followed by two fully connected layers, which gives a vector of 256. The non-spatial view is then concatenated with it before feeding it into another fully connected layer. The last layer has the output size of 33, which corresponds to the number of actions. All layers are followed by Rectified Linear Unit (ReLU) activation function. 




During learning, to encourage exploration we implement epsilon-greedy policies with epsilon piece-wise linear decay over time. Each agent updates its policy given a stored batch (``replay buffer'') of experienced transitions.
Learning agents are ``independent'' of one another and each regard others as part of the environment. From the perspective of a player, the learning of other players shows up as a non-stationary environment. Each individual agent's learning depends only on the other agent's learning via the (slowly) changing distribution of experience it generates. Codes will be made public.

\section{Results}
We will discuss the results separately for the two major components of the proposed framework: detecting and boycotting.

\subsection{Detector}
For Experiment 1 and Experiment 3, a rule-based detector is sufficient to achieve 100\% accuracy. Thus, here we report the result from the trained detector in Experiment 2. We extracted 10000 trajectories of trained agents, half of them are extracted from agents that obey the regulation where the other half are taken from agents that disobey the regulation. We preprocess those trajectories into sequence of rewards and randomly select 20\% of them as the testing set. Using a 4-layer fully connected neural network as the classifier, we ran prediction on the testing set after 100 epochs of training. All layers use Rectified Linear Unit (ReLU) activation function except the last output layer which uses Sigmoid activation function. Results are shown in Figure \ref{fig:exp-detector}. We can see that in this experiment the behavior of agents that do not implement the diminishing reward can be detected with high accuracy, and using longer sequence of observations yields better performance.

\begin{figure}[ht]
   \centering
    \includegraphics[width=.7\linewidth]{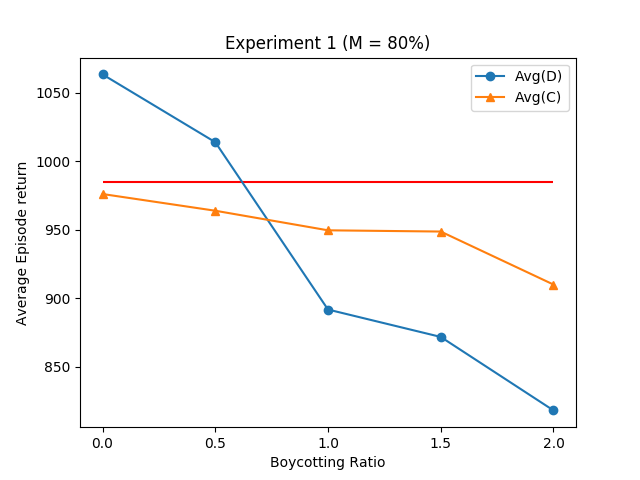}
    \captionof{figure}{Illustration of Table \ref{tab:exp1}. The horizontal red line denotes the average episode return the \textit{Defective} agents will obtain if it behaves compliantly instead.}
    \label{fig:exp1}
    \vspace{-2mm}

\end{figure}
\begin{figure}[ht]
    \centering
    \includegraphics[width=.7\linewidth]{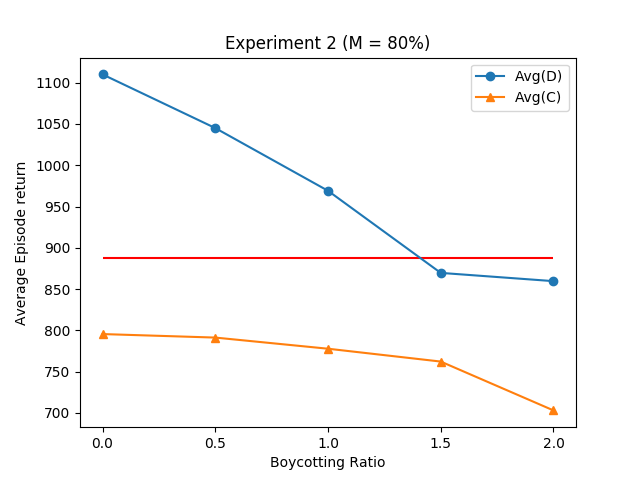}
    \captionof{figure}{Illustration 
3
 of Table \ref{tab:exp2}. The horizontal red line denotes the average episode return the \textit{Defective} agents will obtain if it behaves compliantly instead.}
    \label{fig:exp2}
     \vspace{-2mm}

\end{figure}

\begin{table}[ht]
\centering

\resizebox{.6\columnwidth}{!}{%
\begin{tabular}{ccc}
\cline{1-2}
\multicolumn{1}{|c|}{}                 & \multicolumn{1}{c|}{Avg(C)} &                             \\ \cline{1-2}
\multicolumn{1}{|c|}{All Compliant}  & \multicolumn{1}{c|}{984.7} &                             \\ \cline{1-2}
                                       &                             &                             \\ \hline
\multicolumn{1}{|c|}{Boycotting Ratio} & \multicolumn{1}{c|}{Avg(C)} & \multicolumn{1}{c|}{Avg(D)} \\ \hline
\multicolumn{1}{|c|}{0.0 (original)}   & \multicolumn{1}{c|}{976.0} & \multicolumn{1}{c|}{1063.3} \\ \hline
\multicolumn{1}{|c|}{0.5}              & \multicolumn{1}{c|}{963.8} & \multicolumn{1}{c|}{1013.9} \\ \hline
\multicolumn{1}{|c|}{1.0}              & \multicolumn{1}{c|}{949.5} & \multicolumn{1}{c|}{891.7} \\ \hline
\multicolumn{1}{|c|}{1.5}              & \multicolumn{1}{c|}{948.6} & \multicolumn{1}{c|}{871.7} \\ \hline
\multicolumn{1}{|c|}{2.0}              & \multicolumn{1}{c|}{909.9} & \multicolumn{1}{c|}{818.0} \\ \hline
\end{tabular}

}

\vspace{2mm}
      \captionof{table}[b]{Result of Experiment 1. Avg(D) denotes the average episode return of \textit{Defective} agents while Avg(C) denotes the average episode return of \textit{Compliant} agents.}
      \label{tab:exp1}
\end{table}

\begin{table}[ht]
\centering

\resizebox{.6\columnwidth}{!}{%
\begin{tabular}{ccc}
\hline
\multicolumn{1}{|c|}{}                 & \multicolumn{1}{c|}{Weak}    & \multicolumn{1}{c|}{Strong} \\ \hline
\multicolumn{1}{|c|}{All Compliant}  & \multicolumn{1}{c|}{796.2}  & \multicolumn{1}{c|}{887.2} \\ \hline
                                       &                              &                             \\ \hline
\multicolumn{1}{|c|}{Boycotting Ratio} & \multicolumn{1}{c|}{Avg(C)}  & \multicolumn{1}{c|}{Avg(D)} \\ \hline
\multicolumn{1}{|c|}{0.0 (original)}   & \multicolumn{1}{c|}{795.5}  & \multicolumn{1}{c|}{1110.0} \\ \hline
\multicolumn{1}{|c|}{0.5}              & \multicolumn{1}{c|}{791.2}  & \multicolumn{1}{c|}{1045.0} \\ \hline
\multicolumn{1}{|c|}{1.0}              & \multicolumn{1}{c|}{777.8}  & \multicolumn{1}{c|}{969.0} \\ \hline
\multicolumn{1}{|c|}{1.5}              & \multicolumn{1}{c|}{762.2} & \multicolumn{1}{c|}{869.6} \\ \hline
\multicolumn{1}{|c|}{2.0}              & \multicolumn{1}{c|}{703.1} & \multicolumn{1}{c|}{859.6} \\ \hline
\end{tabular}

}

\vspace{2mm}
    \captionof{table}[b]{Result of Experiment 2. Avg(D) denotes the average episode return of \textit{Defective} agents while Avg(C) denotes the average episode return of \textit{Compliant} agents.}
    \label{tab:exp2}

\end{table}

\subsection{Boycotting Reward Shaping}
The results of Experiments 1 and 2 are shown in Figure \ref{fig:exp1} and Table \ref{tab:exp1} as well as Figure \ref{fig:exp2} and Table \ref{tab:exp2} respectively. Episode return is calculated by counting the number of apples (shared resource) agents collect in an episode. From the tables, we can see that if we assume 80\% (4 out of 5) of the agents are \textit{Compliant}, our goal as stated in Equation \ref{eq:goal} can be fulfilled by setting the \textit{Boycotting Ratio} $B$ to 1.0 or higher for Experiment 1 and 1.5 or higher for Experiment 2. The goal is achieved when the blue line goes below the red line in Figure both \ref{fig:exp1} and \ref{fig:exp2}, which means gaming on the system through violating the regulation can result in worse reward. Below we further describe some important observations.

First, we can observe that higher \textit{Boycotting Ratio} leads to lower return of both \textit{Defective} and \textit{Compliant} agents. This is reasonable because a higher \textit{Boycotting Ratio} means that \textit{Compliant} agents are more encouraged to boycott the \textit{Defective} agents or consume resources that are more likely of \textit{Defective} agents' interest. As result, the \textit{Defective} agent gains lower return, and the \textit{Compliant} agents also gain lower return because their objective is ``deviated'' from maximizing their own rewards. We can think of the boycotting ratio as a measure of the boycotting power we distilled from compliant agents. Higher boycotting ratio is not necessarily better since it leads to less reward for those compliant agents’ reward (Fig. \ref{fig:exp1} and \ref{fig:exp2}.). This can cause the percentage of compliant agents to decrease. To further illustrate this, we can draw an analogy between the boycotting ratio and tax rate. The higher the tax rate is, the government gets more money but more people may try to evade taxes. In short, the boycotting ratio and the percentage of compliant agents forms a dynamic equilibrium. We left this topic as future work since this can be an independent line of research.

Note that all agents are equally capable (meaning that two agents will have the same expected return if they are both \textit{Compliant} or both \textit{Defective}) in Experiment 1 but not in Experiment 2. In Experiment 2, we deliberately make agents boycott the agents who are not only \text{Defective} but also stronger inherently.
The first row in Table \ref{tab:exp2} shows that even if all agents implement diminishing rewards, stronger agents still get considerably more reward than the weaker ones. That is the essence of diminishing reward shaping \cite{sun2018designing} since it still maintains non-homogeneous equality (Stronger agents can still obtain more resources than the weaker ones). However, the gap between them becomes much larger (from 81.0 to 314.5) if the stronger agent opts to be \textit{Defective}. It reassure the necessity of a framework like ours to discourage cheating, as otherwise the stronger agents will have much higher motivation to not obey the regulation. We can also observe that a higher \textit{Boycotting Ratio} is required in Experiment 2 to successfully boycott the \textit{Defective} agent. That can also be explained by the fact that \textit{Defective} agents in Experiment 2 are inherently stronger. 

The result of Experiment 3 is shown in Table \ref{tab:exp3}. If we view the 8 other agents as part of the dynamic environment, and the two remaining players only have two strategies $\{C, D\}$ to choose from, the result can then be interpreted as the payoff matrix of a general-sum matrix game. We can see how the payoff matrix evolves. Before applying our framework, mutual defection is the \emph{Nash Equilibrium}. After our framework is applied, mutual compliant becomes the new \emph{Nash Equilibrium}. This illustrates that our solution is able to promote compliance. Note that it is reasonable that after applying our method, agents gain less rewards. Recall that our goal is to ensure agents follow regulations that aim to ensure sustainable development of resources. Violating the regulation will surely lead to short term gain of rewards (i.e., 844.2 compared to 728.1) but sacrifices long-term sustainability.

\begin{table}[h]
\centering

\resizebox{.6\columnwidth}{!}{%

\begin{tabular}{lll}
\hline
\multicolumn{1}{|l|}{Before} & \multicolumn{1}{l|}{C}                       & \multicolumn{1}{l|}{D}                      \\ \hline
\multicolumn{1}{|l|}{C}      & \multicolumn{1}{l|}{728.1, 728.1}          & \multicolumn{1}{l|}{677.4, 935.0}         \\ \hline
\multicolumn{1}{|l|}{D}      & \multicolumn{1}{l|}{935.0, 677.4}          & \multicolumn{1}{l|}{\textbf{844.2,844.2}} \\ \hline
                             &                                              &                                             \\ \hline
\multicolumn{1}{|l|}{After}  & \multicolumn{1}{l|}{C}                       & \multicolumn{1}{l|}{D}                      \\ \hline
\multicolumn{1}{|l|}{C}      & \multicolumn{1}{l|}{\textbf{728.1, 728.1}} & \multicolumn{1}{l|}{683.1, 481.0}          \\ \hline
\multicolumn{1}{|l|}{D}      & \multicolumn{1}{l|}{481.0, 683.1}           & \multicolumn{1}{l|}{677.8, 677.8}         \\ \hline
\end{tabular}
}
\vspace{2mm}
 \caption{Result of Experiment 3. A strategy profile (cell) is boldfaced if it is a \emph{Nash Equilibrium}.}
 \label{tab:exp3}
 \vspace{-5mm}
\end{table}

\section{Conclusions}
In this paper, we first propose the problem \textbf{Regulation Enforcement} from both RL and game theory perspective. We also present a solution to the problem which aims to eliminate the incentive of agents violating regulations in order to gain more rewards in multi-agent reinforcement learning scenarios. Our solution involves two major components: a detector to identify \textit{Defective} agents and a new regulation that states a boycotting strategy. We demonstrate the effectiveness of the method under two different scenarios - \emph{Replenishing Resource Management Dilemma} and \emph{Diminishing Reward Shaping Enforcement}. We also show how the empirical payoff matrices evolves after applying our method, using empirical game-theoretic analysis. The proposed method ``transfers'' the Nash Equilibrium from mutual defective to mutual compliant.

We expect our paper to ignite more research along this line since there are many future challenges remained. Below, we listed some major challenges that worth more investigation: can we provide a theoretical analysis for the lower bound $M$ (the percentage of compliant agents)? How can we quantify impact of detection errors on the effectiveness of the boycott mechanism?
How to train effective detectors efficiently in the context of online multi-agent learning? How does the  boycotting power distilled from compliant agents relate with the percentage of agents that remain compliant if we consider emotions as a factor? Can we model emotions explicitly in our simulation and how?

\newpage
\bibliographystyle{named}
\bibliography{ijcai19}

\end{document}